\documentclass[10pt,conference,a4paper]{IEEEtran}

\usepackage{fancyhdr} 
\usepackage[USenglish,american]{babel}
\usepackage{epsfig,graphics,subfigure,graphicx,latexsym,longtable,amsmath,amscd,latexsym,amssymb,mathrsfs,syntonly,eucal}
\usepackage{multirow}
\usepackage[usenames]{color}
\usepackage[T1]{fontenc}
\usepackage{bm,cite}
\usepackage{amsbsy}
\usepackage{latexsym}
\usepackage{wasysym}
\usepackage{placeins}
\usepackage[lined,boxed,commentsnumbered]{algorithm2e}
\usepackage{lipsum}
\usepackage{url}
\usepackage{array}
\usepackage{tabu}
\usepackage[skip=4pt,font=scriptsize]{caption}
\SetKwInput{KwInput}{Input}
\SetKwInput{KwOutput}{Output}
\usepackage{booktabs,siunitx}
\usepackage{geometry}
\usepackage{bm,cite}
\usepackage{amsmath}
\usepackage{amsbsy}
\usepackage{latexsym}
\usepackage{amssymb}
\usepackage{wasysym}
\usepackage{mathtools}

\DeclareMathAlphabet{\mathbit}{OML}{cmr}{bx}{it}
\DeclareMathAlphabet{\mathsf}{OT1}{cmss}{m}{n}
\DeclareMathAlphabet{\mathTXf}{OT1}{cmss}{bx}{it}

\DeclareMathOperator{\diag}{diag}

\DeclareMathOperator{\Transpose}{T}

\DeclareMathOperator*{\argmax}{arg\ max}

\DeclareMathAlphabet{\mathpzc}{OT1}{pzc}{m}{it}

\newcommand{\TA}{{\text{TA}}}

\newcommand{\Tr}{{\Transpose}}

\newcommand{\He}{{{\text{H}}}}

\graphicspath{{./figures/}}

\title{Novel Round Trip Time Estimation in 5G NR}

\author{
\IEEEauthorblockN{Rakesh Mundlamuri\IEEEauthorrefmark{1}\IEEEauthorrefmark{2}, Rajeev Gangula\IEEEauthorrefmark{2}, Florian Kaltenberger\IEEEauthorrefmark{1}\IEEEauthorrefmark{2}, and Raymond Knopp\IEEEauthorrefmark{1} 
}
\IEEEauthorblockN{\IEEEauthorrefmark{1}Communication Systems Department,
EURECOM, Biot, France
}

\IEEEauthorblockN{\IEEEauthorrefmark{2}Institute for the Wireless Internet of Things, Northeastern University, Boston, USA 
}
}

\begin{document}
\maketitle

\begin{abstract}
The fifth generation new radio (5G NR) technology is expected to fulfill reliable and accurate positioning requirements of industry use cases, such as autonomous robots, connected vehicles, and future factories. Starting from Third Generation Partnership Project (3GPP) Release-16, several enhanced positioning solutions are featured in the 5G standards, including the multi-cell round trip time (multi-RTT) method. This work presents a novel framework to estimate the round-trip time (RTT) between a user equipment (UE) and a base station (gNB) in 5G NR. Unlike the existing scheme in the standards, RTT can be estimated without the need to send timing measurements from both the gNB and UE to a central node. The proposed method relies on obtaining multiple coherent uplink wide-band channel measurements at the gNB by circumventing the timing advance control loops and the clock drift. The performance is evaluated through experiments leveraging a real world 5G testbed based on OpenAirInterface (OAI). Under a moderate system bandwidth of 40MHz, the experimental results show meter level range accuracy even in low signal-to-noise ratio (SNR) conditions.
\end{abstract} 
\section{Introduction} \label{sec:intro}

Precise and reliable location awareness plays a crucial role in the development of intelligent, autonomous, and connected systems \cite{KuSamSab_18,EskWuChu_21,9921342}. Besides offering cellular connectivity, owing to the usage of large bandwidth and massive antenna arrays, 5G new radio (5G NR) is expected to provide highly accurate positioning capability \cite{DwivediEtal_21,italiano2024tutorial,KeaSalMik_19}. While the 5G-based solution can complement the Global Positioning System (GPS) in outdoor scenarios, it provides a viable alternative in low-visibility, indoor, and GPS-denied environments. In addition, location awareness of user devices is known to greatly improve cellular network performance \cite{Dirk_12,DiMupRauSlock_14,HamPetGabor_22}.

From Release-16 onwards, the Third Generation Partnership Project (3GPP) has started defining the 5G NR target positioning accuracy for various scenarios. New positioning signals, measurement procedures, and architecture are accommodated \cite{3gpp2018nr_38_855,3gpp2018nr_38_455_16,3gpp2018nr_38_857,3gpp2018nr_38_455_17,WanHuaYiCh_23}. The positioning methods generally involve extracting time, angle, and signal strength information at the base station (gNB) and user (UE) from the received reference radio signals. Timing-based methods include enhanced cell ID (E-CID), multi-cell or single-cell round trip Time (RTT), downlink time difference of arrival (DL-TDoA), and uplink time difference of arrival (UL-TDoA). Techniques such as downlink angle of departure (DL-AoD) and uplink angle of arrival (UL-AoA) rely on multiple antennas to obtain the angular information from the measurements \cite{DwivediEtal_21,italiano2024tutorial}.

In DL-TDoA and UL-TDoA methods, the time difference of arrival from multiple gNBs is used in localization. This method requires the participating gNBs to be tightly synchronized. In the multi-cell RTT method, absolute time-of-arrival (ToA) is calculated at multiple gNBs. Unlike the DL-TDoA and UL-TDoA methods, synchronization among gNBs is not needed. Even when only one multi-antenna gNB is available, range estimation from RTT can be combined with the UL-AoA measurements to localize the UE.

There are mainly two approaches for obtaining RTT in 5G NR. Once the UE is synchronized in downlink (DL), the RTT in the form of timing advance (TA) can be estimated from a received random access channel (RACH) preamble during the random access (RA) procedure. Even though this appears to be a straightforward approach, it suffers from low accuracy due to the limited bandwidth of the RACH. Moreover, the UE performs the RA procedure only during the initial access or when the uplink (UL) synchronization is lost. In a more dedicated method, the DL positioning reference signal (PRS) and UL sounding reference signal (SRS) are used to estimate the receive-transmit timing difference at the UE and gNB. These measurements are then exchanged to a central node to calculate the RTT. The accuracy of this method comes at the expense of using more DL and UL resources \cite{DwivediEtal_21,italiano2024tutorial}.

Accurate RTT estimation is possible if the timing measurements from RACH can be augmented with wideband UL channel measurements, for example, from SRS. Few works have considered the problem of estimating RTT based on UL channel measurements at the gNB in real-world scenarios using open-source 4G/5G testbeds \cite{BlAlLuZhe_19,LiChuWaZh_22}. The work in \cite{BlAlLuZhe_19} is restricted to a scenario where there is no TA correction sent to the UE by the gNB after the initial random access, i.e., the UE is very close to the gNB. An improved scheme that overcomes this restriction has been proposed in \cite{LiChuWaZh_22}. However, both works ignore the effect of clock drift in the system. Indeed, large fluctuations in RTT estimates caused by these factors are reported in \cite{LiChuWaZh_22}. Moreover, RTT can only be obtained during initial access.

The inability to exploit multiple UL SRS measurements coherently in the RTT estimation stems from a) inherent timing control loops in 5G NR and b) clock drift. The timing control loops in 5G NR include UL and DL timing control. UL timing control is a continuous process in which gNB sends TA commands to the UE to adjust its UL transmission timing. This procedure is crucial for maintaining UL frame alignment with the gNB. On the other hand, in DL timing control, the UE experiences DL reception timing drift due to clock drift, and it corrects this drift based on DL reference signals and is implementation specific. These timing control loops and the clock drift lead to the variability in delay estimated from SRS measurements obtained in different time slots. Therefore, even in a scenario where the UE is static, and the gNB has access to multiple SRS measurements, they cannot be used jointly to estimate the RTT. However, it is well known that multiple measurements tend to improve the estimation performance in low signal-to-noise ratio (SNR) conditions.

In this work, we propose a novel framework to estimate the RTT based on multiple coherent SRS measurements in 5G NR. This approach tremendously improves the RTT estimation accuracy in the low SNR regime. To the best of the author's knowledge, accurate RTT estimation without the need for dedicated DL PRS resources is not possible in 5G NR.
Specifically, the contributions of this paper are: 
\begin{itemize}
    \item We propose a simple enhancement to the 5G NR signaling scheme capable of obtaining a sequence of similar UL SRS measurements.
    \item A matched-filter solution is proposed to estimate the RTT jointly from the 
    collected measurements.
    \item The proposed method can obtain the RTT even when the 5G UE is in a radio resource control (RRC) inactive state.
    \item The complete solution is experimentally validated with a real-word 5G 
    testbed based on the OpenAirInterface (OAI)\cite{KalaloAbhiLuh_20}.
\end{itemize}

The rest of the paper is organized as follows: Section~\ref{sec:backg} reviews the 5G concepts related to positioning. In Section~\ref{sec:signaling}, we describe the RTT estimation framework and the proposed signaling enhancements to 3GPP. Section~\ref{sec:ToAesti} presents the estimation process and algorithms. Section~\ref{sec:Expsetup} outlines the detailed experimental setup, and the evaluation results are presented in Section~\ref{sec:Results}. Finally, Section~\ref{sec:conclusion} concludes the paper.

\section{Background}\label{sec:backg}
This section provides an overview of the 5G positioning related concepts and radio resource control (RRC) connectivity states of a user (UE).

\subsection{5G Synchronization}\label{sec:5gsync}

Synchronization between base station (gNB) and UE is essential for reliable communication as well as positioning. The 5G synchronization process consists of downlink (DL) and uplink (UL) synchronization. The UE can detect symbol and frame boundary during the DL synchronization. The gNB periodically transmits a synchronization signal block (SSB) in the DL. The SSB consists of a primary synchronization signal (PSS), a secondary synchronization signal (SSS), and a physical broadcast channel (PBCH) with its associated demodulation reference signal (DMRS). Once DL is synchronized, the UE extracts configuration parameters by decoding the master information block (MIB) from the PBCH and later the system information block (SIB) from the physical downlink shared channel (PDSCH). These parameters provide the necessary information to perform UL synchronization.

The UL synchronization enables UE to determine the exact time to send the UL data. Since a gNB serves multiple UEs located across the cell, the UL transmission times of the UEs need to be adjusted such that their reception is aligned with the gNB's UL reception. This is achieved with a random access (RA) procedure. The UL synchronization is as follows:
\begin{itemize}
    \item Either the UE initiates (in the case of initial access), or the gNB orders the UE (in case of loss of UL synchronization) to initiate the RA procedure through the physical random access channel (PRACH). While it is contention based RA in the former scenario, in the latter, it can be contention free, i.e., the gNB may configure the UE with a dedicated PRACH preamble.
    \item Based on the delay estimated from the PRACH, the gNB can measure a coarse/quantized round-trip time (RTT). This coarse/quantized version of RTT coined as timing advance (TA), is then sent to the UE via the random access response (RAR).
    \item Although the initial TA is sent via the RAR, gNB can periodically send updated UL timing corrections to the UE via TA commands. Through these TA commands, gNB can maintain the UL synchronization in case of UE mobility or clock drift.
\end{itemize}

\begin{figure}[t]
\centerline{\includegraphics[width=3in]{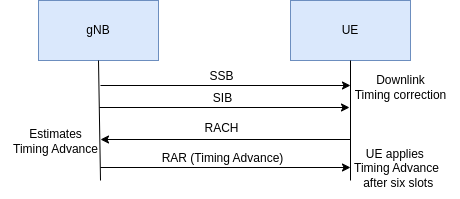}}
\caption{5G synchronization.}
\label{fig:sync_call_flow}
\vspace{-3mm}
\end{figure}

\begin{figure}[t]
\centerline{\includegraphics[width=3in]{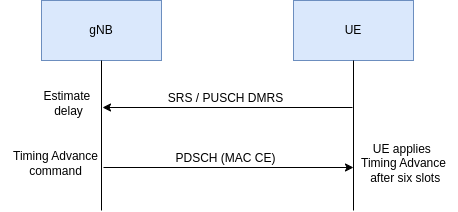}}
\caption{UE UL timing correction with TA commands.}
\label{fig:ta_command}
\vspace{-6mm}
\end{figure}

The signaling procedure of RA and TA updates is depicted in
Figures~\ref{fig:sync_call_flow} and~\ref{fig:ta_command}.

\subsection{Reference Signals}\label{sec:refsig}
The reference signals or pilots are used in channel estimation,
parameter estimation, and positioning. The widely used reference signals for positioning in 5G NR are as follows:
\begin{itemize}
    \item The Sounding Reference Signal (SRS) is a wide-band reference signal transmitted by the UE in the UL. SRS is generated using the Zadof-Chu sequence and has good auto and cross-correlation properties. Dedicated SRS for positioning is introduced in 3GPP Release-11. Although SRS for positioning and communication have a lot of commonalities, they can be configured separately \cite{DwivediEtal_21}.
    
\item 3GPP has introduced the PRS, especially for localization purposes in the DL\cite{3gpp2018nr_38_211}. These reference signals were introduced in 4G and extended to 5G with better resolution and accuracy. PRS is generated using QPSK modulated 31-length gold sequence. PRS can be flexibly arranged in any number of PRBs in the frequency domain. In the time domain, the PRS resources for positioning can span \{2,4,6,12\} consecutive OFDM symbols. 
\end{itemize}

\subsection{5G RRC states}\label{sec:uerrc}

In 5G standalone mode, three possible radio resource control (RRC) states are defined for a UE: NR\_RRC\_IDLE, NR\_RRC\_INACTIVE, and NR\_RRC\_CONNECTED. While the NR\_RRC\_CONNECTED indicates an active data transmission/reception, the UE in NR\_RRC\_IDLE and NR\_RRC\_INACTIVE stays asleep most of the time and periodically wakes up and looks for paging messages to look for active data (e.g., incoming voice call or data) and switches to NR\_RRC\_CONNECTED state. Contrary to the NR\_RRC\_IDLE mode, in NR\_RRC\_INACTIVE state, the gNB stores the UE context (e.g., RRC configuration) for periodic transmissions. It is to be noted that until 3GPP Release-16, majority of the current 5G positioning techniques that offer good accuracy work only when the UE is NR\_RRC\_CONNECTED. 3GPP Release-17 introduces NR\_RRC\_INACTIVE based positioning for low power high accuracy positioning (LPHAP)\cite{LPHAP} for low power devices. The design requirements for the LPHAP include low power consumption, low complexity, low signaling overhead, and timing alignment to avoid interference with other UEs\cite{rel_17_design}. These design requirements make the proposed RTT estimation framework well-suitable for NR\_RRC\_INACTIVE based positioning.
Next, we describe the signaling enhancements needed in 3GPP to enable the proposed framework.





\section{Signaling Enhancements}\label{sec:signaling}
The proposed signaling scheme aims to obtain multiple wide-band uplink (UL) sounding reference signal (SRS) measurements along with the RACH
at the base station (gNB). We leverage on the existing physical downlink control channel (PDCCH) order signaling mechanism in 5G NR to obtain these measurements. When an RTT request is made, the gNB sends an enhanced downlink control information (DCI) that includes information fields related to positioning to the user (UE) as a DCI Format via PDCCH. Moreover, the DCI Format should consider both the NR\_RRC\_INACTIVE and NR\_RRC\_CONNECTED states of a UE. 

\begin{figure}[t]
\centerline{\includegraphics[width=3in]{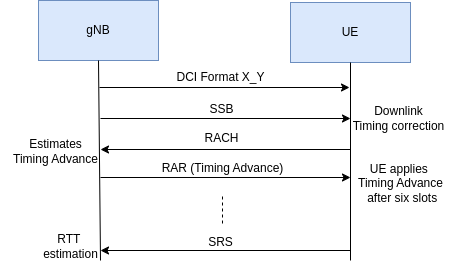}}
\caption{Proposed signaling mechanism for RTT estimation.}
\label{fig:pdcch_order}
\vspace{-5mm}
\end{figure}

The proposed 3GPP like signaling scheme is depicted in Figure~\ref{fig:pdcch_order}. A new DCI Format, termed as DCI Format X\_Y, is used to signal the UE for positioning.
Once the UE decodes the DCI Format X\_Y, it adjusts or updates its downlink (DL) synchronization based on the synchronization signal block (SSB)\footnote{Note that UE can update its DL synchronization by other means too }. The UE then triggers a contention free RACH using the dedicated preamble mentioned in the DCI Format X\_Y. The gNB estimates the timing advance (TA) from the RACH preamble and sends it to the UE via RAR. The UE applies the TA after six slots (from current standards) and then transmits the SRS, as shown in Figure~\ref{fig:pdcch_order}. RTT can be calculated based on the TA and the SRS channel estimates. This procedure is repeated several times to obtain multiple measurements. Note that the UE aligns its UL timing when sending the SRS using the TA value received in the RAR.

The DCI Format X\_Y\cite{rtt_patent} includes the following fields: fullI-RNTI or shortI-RNTI, Random Access Preamble index, UL/SUL indicator, SS/PBCH index, PRACH Mask index, and SRS request as shown in Table.~\ref{tab:dci}. In the proposed signaling mechanism,
\begin{itemize}
    \item when UE is NR\_RRC\_INACTIVE state, the DCI Format X\_Y is scrambled using P-RNTI, 
    \item in NR\_RRC\_CONNECTED state, the DCI Format X\_Y is scrambled using C-RNTI. In this case, the field fullI-RNTI or shortI-RNTI is set to 0. 
\end{itemize}
Alternatively, it is possible to use the existing DCI Formats in the current 3GPP standard with minor modifications\cite{rtt_patent}. This can be achieved by adding an SRS request field to the DCI Format 1\_0 for RAN paging in the NR\_RRC\_INACTIVE state and the DCI Format 1\_0 for PDCCH order in the NR\_RRC\_CONNECTED state. However, the DCI Format 1\_0 for RAN paging initiates a contention based RACH procedure, resulting in signaling overhead. 

\begin{figure}[t]
\centerline{\includegraphics[width=3in]{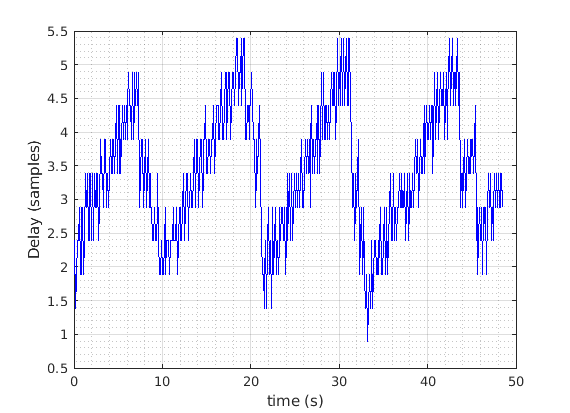}}
\caption{Effect of clock drift on RTT in commercial UE.}
\label{fig:ramp}
\vspace{-5mm}
\end{figure}

Furthermore, DL timing correction is crucial for positioning when a DCI is received. This can be observed from the behavior of a 3GPP Release-15 commercial UE (Quectel RM500-GL) as shown in Figure~\ref{fig:ramp}. The plot in Figure~\ref{fig:ramp} indicates the RTT estimated over time by combining TA from RACH and SRS measurements. The signaling mechanism for estimating RTT is similar to the proposed mechanism as depicted in Figure~\ref{fig:pdcch_order}. An existing DCI (PDCCH order) is used to trigger the RACH, and the SRS is scheduled so that the UE transmits the SRS after applying the TA received from RAR. The sawtooth structure observed from the RTT measurements over time is shown in Figure~\ref{fig:ramp}.
In this sawtooth behavior,
\begin{itemize}
    \item The rise in the RTT estimates stems from the clock drift between the gNB and the UE because the UE does not correct its DL timing immediately after receiving the DCI.
    \item The fall occurs when the UE corrects its DL timing.
\end{itemize}
 Currently, commercial UEs using 3GPP Release-15 have an implementation specific timing correction that corrects the DL timing only based on the conformance requirement\cite{3gpp2018nr_38_533} but not when the DCI is received.

Therefore, the proposed DCI Format X\_Y reduces the signaling overhead, maintains DL and UL timing, and uses a common DCI for both RRC states.

\begin{table}[htbp]
\caption{DCI Format X\_Y}
\centering
\begin{tabular*}{\columnwidth}{@{\extracolsep{\fill}}cc}
\toprule
DCI Fields & Number of bits \\
\midrule
fullI-RNTI or shortI-RNTI & 40 or 24 \\
Random Access Preamble index & 6 \\
UL/SUL indicator & 1 \\
SS/PBCH index & 6 \\
PRACH Mask index & 4 \\
SRS request & 2 or 3\\
\bottomrule
\end{tabular*}
\label{tab:dci}
\end{table}

\section{RTT Estimation}\label{sec:ToAesti}
Based on the signal enhancements described in the previous section, the gNB can estimate a coarse RTT, i.e., $\TA$ value estimated from the RACH, and refine it further using SRS measurements. The coarse RTT ${\tau}^r$ can be obtained from the $\TA$ as
\begin{equation}
 {\tau}^r = \frac{\TA\times16\times64}{2^{\mu}}\text{T}_c,
\label{eq:RTT}
\end{equation}
where, $\mu\in\{0,1,2,3,4,5\}$ is the numerology related to the subcarrier spacing $\Delta f=15.2^{\mu}\ \textrm{KHz}$, $\text{T}_c=\frac{1}{(\Delta f_{max}\times K_{max})}$, $\Delta f_{max}=480\ \text{KHz}$ is the maximum possible sub-carrier spacing and $K_{max}=4096$ is the maximum possible FFT size in NR \cite{3gpp2018nr_38_211}. Next, we describe the SRS channel estimation.

\subsection{SRS channel estimation}
We consider a 5G NR system with a single antenna UE and gNB in our experiments. The received sounding reference signal (SRS) at the gNB on the $k$-th, $k\in[0,K-1]$, subcarrier
\begin{equation}
    y[k] = h[k]s[k] + n[k],
\end{equation}
where, $k\in\{0,1,\dots,{K}{-}1\}$, $K$ is the FFT size, $h[k]$ represents the baseband propagation channel, $s[k]$ represents the SRS pilot symbol and $n[k]\sim\mathcal{N}(0, \sigma^2)$ is additive white gaussian noise. We consider a line-of-sight (LoS) channel with no multi-path components. The UL channel at a $k$-{th} subcarrier is modeled as
\begin{equation}
    h[k] = \alpha e^{-j2\pi k\Delta f \tau},
\end{equation}
where, $\alpha =\alpha^{\prime}e^{-j2\pi f_c \tau}$ is the complex channel gain of the path, $\alpha^{\prime}$ is the path-loss, $\Delta f$ is the subcarrier spacing, $f_c$ is the centre frequency.

A least-square estimate of the channel is given by
\begin{equation}
    \hat{h}[k] = s[k]^*y[k],
\end{equation}
where, $(.)^*$ denotes the conjugate operator, and $s[k]$ is the known SRS symbol at the $k$-th subcarrier.

The channel estimate $\hat{h}[k]$ is represented in the vector form as
\begin{align}
    \bm{\hat{h}} = \left[\hat{h} [0], \hat{h}[1] ,\dots,\hat{h}[{K}{-}1]\right]^{\Tr}.
\end{align}

We now further refine the RTT estimate using the SRS channel estimates. 

\subsection{Matched Filter} \label{sec:mf}
By using the proposed signaling scheme in Section~\ref{sec:signaling}, we can obtain multiple coarse RTT's and SRS channel estimates. The $m$-th, $m \in [1,M]$ coarse RTT obtained from RACH~\eqref{eq:RTT} and the channel estimate is denoted by $\tau_m^r$ and $\bm{\hat{h}}_m \in\mathbb{C}^{{K}\times 1}$ respectively. Each channel estimate $\bm{\hat{h}}_m$ is shifted by the respective coarse RTT $\tau_m^r$ to maintain the coherency among the channel estimates. A matched filter estimate of the refined RTT $\hat{\tau}$ is given by
\begin{equation}
    \hat{\tau} = \argmax_{{\tau}} \frac{1}{{M}}\sum^{{M}}_{m=1}|\bm{v}(\tau)^{\He}\mathbf{T}(\tau_m^{r})\bm{\hat{h}}_m|^2,
\end{equation}
where, $\bm{v}(\tau)=
[1,e^{-j2\pi \Delta f \tau},\ldots,e^{-j2\pi (K-1)\Delta f \tau}]^\Tr$ and $\mathbf{T}(\tau_m^{r})=
\diag(1,e^{-j2\pi \Delta f \tau_m^{r}},\ldots,e^{-j2\pi (K-1)\Delta f \tau_m^{r}})$.

Further, the distance estimate ($\hat{d}$) between the gNB and the UE can be obtained from the estimated RTT ($\hat{\tau}$) as
\begin{equation}
    \hat{d} = \frac{\hat{\tau}\text{c}}{2},
\end{equation}
where, c is the speed of light. 

The performance of the proposed algorithm is evaluated using the following experimental setup.

\begin{figure*}
    \centering    \includegraphics[width=\textwidth]{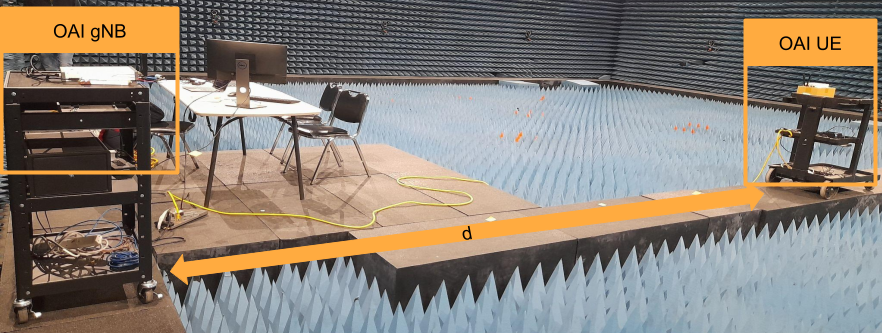}
    \caption{Experimental setup in the anechoic chamber.}
    \label{fig:chamber}
    \vspace{-5mm}
\end{figure*}

\section{Experimental Setup}\label{sec:Expsetup}

We consider a scenario with a single antenna gNB and a UE having line-of-sight (LoS) condition. We leverage on the OpenAirInterface (OAI) 5G NR protocol
stack \cite{KalaloAbhiLuh_20} and USRP B210 software-defined radio cards to build the gNB and UE. Additionally, 
SC2430 NR signal conditioning module is used as an external RF front-end at the gNB \cite{sc_mod}. The measurements were taken in an anechoic chamber at the Northeastern University Burlington campus, as shown in Figure~\ref{fig:chamber}.

\begin{figure}[t]
\centerline{\includegraphics[width=3in]{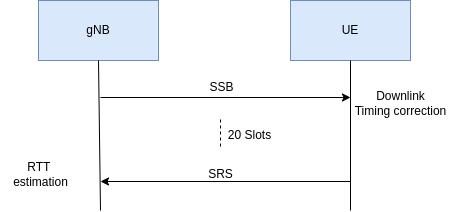}}
\caption{RTT implementation in OAI phy-test mode.}
\label{fig:phy_test_oai}
\vspace{-3mm}
\end{figure}

Since the signaling using the DCI Format X\_Y is not implemented yet, we simplify the proposed RTT signaling scheme depicted in Figure~\ref{fig:pdcch_order} of Section~\ref{sec:signaling} to the signaling scheme as shown in Figure~\ref{fig:phy_test_oai}. Note that this simplification does not result in any loss in terms of the functional behavior of the proposed RTT estimation algorithm for the following reasons:
\begin{itemize}
    \item The distance between the gNB and UE is always within the resolution of the RACH based timing advance (TA) during the experiment. For the used 5G NR configuration, the RACH TA resolution
is 39.0625 meters.
\item The delay between receiving the DCI Format X\_Y and sending the SRS in the proposed framework (Figure~\ref{fig:pdcch_order}) is emulated by a 20 slot offset between SSB reception and SRS transmission at the UE as shown in Figure~\ref{fig:phy_test_oai}.
\item The hardware delays are calibrated and compensated by applying a fixed timing advance at the UE
throughout the experiment. For this, we have used the phy-test mode of the OAI \cite{phy_test}. This mode operates only at the physical layer and abstracts the higher layers.   
\end{itemize}
The 5G NR system parameters used in the experiment are listed in Table~\ref{tab:1}. 

During the experiment, the gNB is static, and the UE is moved in an increment of 1 meter from an initial gNB-UE distance of 7 to 11 meters, as shown in Figure~\ref{fig:chamber}. In all measurements, the LoS is maintained between the gNB and UE. Variation in uplink SNR is achieved by changing the USRP transmit (TX) gain. Multiple SRS channel estimates were obtained at each distance, and the data was then stored for further offline analysis. The performance of the RTT estimation scheme based on the collected measurements is presented in the following Section.

\begin{table}[htbp]
\caption{System Parameters}
\centering
\begin{tabular*}{\columnwidth}{@{\extracolsep{\fill}}cc}
\toprule
Parameters & Values \\
\midrule
System bandwidth & 38.16 MHz \\
Subcarrier Spacing ($\Delta f$) & 30 KHz \\
Centre frequency ($f_c$) & 3.69 GHz \\
Sampling rate ($f_s$) & 46.08 MHz \\
FFT size ($K$) & 1536 \\
Cyclic prefix ($N_{cp}$) & 132\\
SSB bandwidth & 7.2 MHz\\
SRS bandwidth & 37.44 MHz\\
SRS comb size ($K_c$) & 2\\
\bottomrule
\end{tabular*}
\label{tab:1}
\vspace{-6mm}
\end{table}

\section{Results}\label{sec:Results}

In this section, we will present the empirical results of the proposed RTT estimation in low and high uplink SNR scenarios. Moreover, the impact of the number of measurements on the two algorithms: the peak detector (PD) and the matched filter (MF) is shown. While the MF approach is outlined in Section~\ref{sec:mf}, the range estimated using PD is given by
\begin{equation}
\hat{d}  = \frac{\text{c}}{2f_sM}\sum_{m=1}^{M} \argmax \left| IDFT \left\{\bm{\hat{h}}_m\right\}\right|,
\end{equation}
where, $\bm{\hat{h}}_m,m\in[1,M]$ is the $m$-th SRS channel estimate measurement, $f_s$ is the sampling rate, 
$IDFT\{.\}$ denotes the Inverse Discrete Fourier Transform, and $c$ is the speed of light.



\begin{figure}[t]
\centerline{\includegraphics[width=3in]{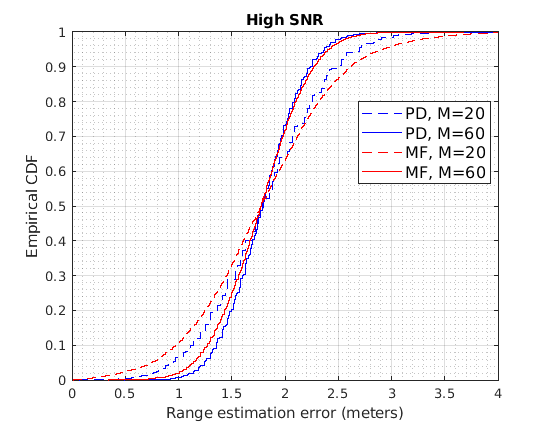}}
\caption{CDF of the range
estimation error.}
\label{fig:high_snr}
\vspace{-5mm}
\end{figure}
\begin{figure}[t]
\centerline{\includegraphics[width=3in]{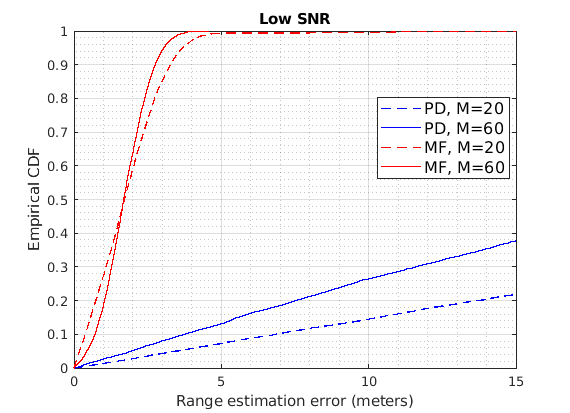}}
\caption{CDF of the range
estimation error.}
\label{fig:low_snr}
\vspace{-5mm}
\end{figure}

The cumulative distribution function (CDF) of the range estimation error for MF and PD algorithms in high and low uplink signal-to-noise ratio (SNR) is shown in Figures~\ref{fig:high_snr} and~\ref{fig:low_snr}. At each SNR, the empirical range CDF is obtained from a total of 25,000 SRS measurements. These measurements are collected based on the signaling procedure depicted in Figure~\ref{fig:phy_test_oai}. As described in Section~\ref{sec:Expsetup}, these 25,000 measurements are obtained by keeping the gNB's position fixed and moving the UE between 7 to 11 meters with a 1 meter increment. At every SNR, 5,000 measurements are collected at each distance between 7 and 11 meters. 

For the high SNR scenario, we fixed the UE USRP TX gain to $89.5$ dB, resulting in an estimated uplink SNR of $25$ dB. For the low SNR scenario, we have reduced the UE TX gain by $50$ dB to $39.5$ dB. While the MF and PD schemes have similar performance when the SNR is high, 
in low SNR scenarios, the MF algorithm significantly outperforms the PD scheme.
In a low SNR scenario, for $M{=}20$, the range estimation error of MF is below 3.25 meters for 90\% of the time. Furthermore, by increasing the number of measurements from $M{=}20$ to $M{=}60$, we see an increase in the estimation performance for both methods.


\section{Conclusion}\label{sec:conclusion}
In this work, we proposed a novel RTT estimation framework using a new DCI Format X\_Y as a signaling mechanism for positioning. The proposed framework does not rely on dedicated DL PRS and works in both NR\_RRC\_INACTIVE and NR\_RRC\_CONNECTED states. It forces the UE to correct its timing after the reception of the DCI, which is currently not possible, as our experiments with COTS UEs have shown. Furthermore, our framework enables the coherent combination of multiple uplink channel measurements and is robust to the clock drift and the inherent timing loops in the 5G system. We have validated the functionality of our proposed framework in real-time using OAI. Our results show that the proposed matched filter algorithm can achieve meter-level accuracy for bandwidth as low as 40MHz, even in low SNR scenarios. 
\bibliographystyle{IEEEtran}
\bibliography{References}

\end{document}